\documentclass[journal]{IEEEtran}

\usepackage{url}
\usepackage{graphicx}
\usepackage{hyperref}
\usepackage{amsmath}
\usepackage{amssymb}
\usepackage{array}
\usepackage{booktabs}
\usepackage{multirow}
\usepackage{makecell}

\begin{document}

\title{Weakly Supervised Anomaly Detection for Chest X-Ray Image}

\author{Haoqi Ni, Ximiao Zhang, Min Xu, Ning Lang, and Xiuzhuang Zhou

\thanks{
This work was supported in part by National Natural Science Foundation of China under Grant 61972046 and Grant 62177034.}

\thanks{Haoqi Ni and Xiuzhuang Zhou are with the School of Artificial Intelligence, Beijing University of Posts and Telecommunications, Beijing 100876, China. E-mail: xiuzhuang.zhou@bupt.edu.cn. (Corresponding author: Xiuzhuang Zhou)}

\thanks{Ximiao Zhang and Min Xu are with the College of Information and Engineering, Capital Normal University, Beijing 100048, China. E-mail: 2211002048@cnu.edu.cn.}

\thanks{Ning Lang is with the Department of Radiology, Peking University Third Hospital, Beijing 100191, China.}

}

\markboth{Journal of \LaTeX\ Class Files, Vol. 14, No. 8, August 2015}
{Shell \MakeLowercase{\textit{et al.}}: Bare Demo of IEEEtran.cls for IEEE Journals}

\maketitle

\begin{abstract}
    Chest X-Ray (CXR) examination is a common method for assessing thoracic diseases in clinical applications. While recent advances in deep learning have enhanced the significance of visual analysis for CXR anomaly detection, current methods often miss key cues in anomaly images crucial for identifying disease regions, as they predominantly rely on unsupervised training with normal images. This letter focuses on a more practical setup in which few-shot anomaly images with only image-level labels are available during training. For this purpose, we propose WSCXR, a weakly supervised anomaly detection framework for CXR. WSCXR firstly constructs sets of normal and anomaly image features respectively. It then refines the anomaly image features by eliminating normal region features through anomaly feature mining, thus fully leveraging the scarce yet crucial features of diseased areas. Additionally, WSCXR employs a linear mixing strategy to augment the anomaly features, facilitating the training of anomaly detector with few-shot anomaly images. Experiments on two CXR datasets demonstrate the effectiveness of our approach.
\end{abstract}

\begin{IEEEkeywords}
 Anomaly detection, weakly supervised, Chest X-Ray image.
\end{IEEEkeywords}

\IEEEpeerreviewmaketitle

\section{Introduction}

\IEEEPARstart{C}{hest} radiography is the most common medical imaging examination. Due to its low cost and clear imaging, it has been widely used and plays a crucial role in the early screening and diagnosis of many kinds of diseases. The objective of anomaly detection in Chest X-Ray (CXR) is to automatically identify whether a given CXR image contains lesion regions, serving as a crucial tool to assist medical professionals in making decisions. Recently, due to the rapid development in deep learning technology, there have been significant advancements in medical image anomaly detection methods \cite{tian2021constrained,naval2021implicit,zhang2022multi,pinaya2022fast,wyatt2022anoddpm,xiang2023squid}, enabling the automation of anomaly classification in medical images. Influenced by the popular unsupervised paradigm of industrial image anomaly detection \cite{tao2022deep,liu2023deep,bergmann2019mvtec,zou2022spot,liang2023omni,huang2022self}, current medical image anomaly detection methods \cite{pinaya2022fast,wyatt2022anoddpm,xiang2023squid,bozorgtabar2023amae,wolleb2022diffusion,bercea2023reversing} solely use normal images for model training, considering regions deviating from the distribution of normal images as disease regions. However, unlike the extremely rare occurrence of anomalies in industrial production, various diseases typically have a relatively high incidence rate. This means that in practice, a limited number of typical lesion images are usually available. Furthermore, compared to the diversity of anomaly regions in industrial images, disease regions in CXR images often exhibit certain regularities. Therefore, using few-shot real lesion CXR images to guide the training of medical anomaly detection models can effectively leverage the scarce yet crucial cues within anomaly images, significantly enhancing anomaly detection performance.

In this letter, we propose a more practical setting for medical image anomaly detection, namely, weakly supervised anomaly detection. In the training process, a substantial number of normal images and few-shot actual anomaly images are available, with only image-level labels. This setup avoids the need for expensive pixel-level annotation. For this purpose, we introduce WSCXR, a weakly supervised anomaly detection framework for CXR. WSCXR can effectively leverage medical cues from few-shot real anomalous images for anomaly detection, thereby improving the model's anomaly detection performance. WSCXR firstly extracts patch features from normal images and anomaly images to construct the sets of normal image features and anomaly image features respectively. Since anomaly images contain a significant number of normal regions, we propose anomaly feature mining to remove normal features from the set of anomaly image features, retaining only the patch features corresponding to the lesion areas for anomaly detection. Due to the scarcity of anomaly features, we further propose a linear mixing strategy to augment the set of anomaly features. The augmented set of anomaly features enables the learned anomaly detector to have a compact decision boundary, thereby enhancing the model’s generalization capability. The most closely related work to ours is \cite{cai2022dual}, which divides the training data into a normal image set and an unlabeled image set. We simplified the setting described in \cite{cai2022dual} by considering the anomaly image set as naturally obtained when partitioning the normal image set, and more importantly, we do not utilize the unlabeled image set for model training. The main contributions of this work can be outlined below:

\begin{figure*}[t]
  \centering
   \includegraphics[width=\linewidth]{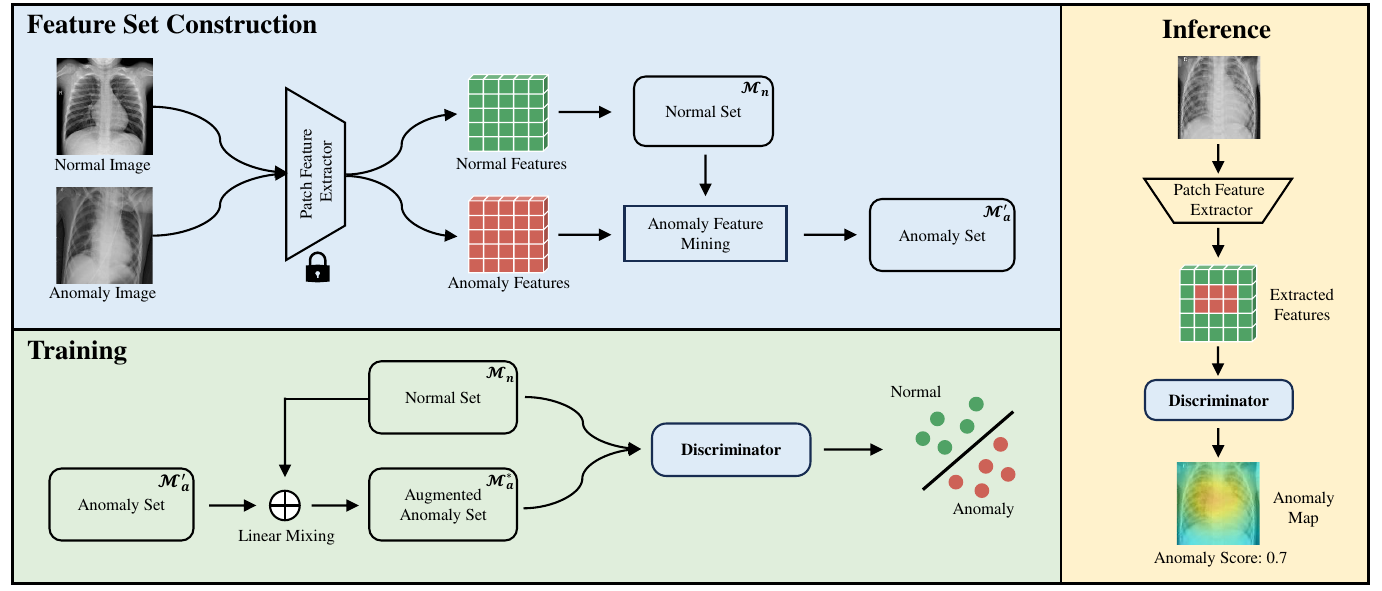}
   \caption{The pipeline of WSCXR. WSCXR extracts patch features from normal image to construct the normal set $\mathcal{M}_{n}$, and then performs anomaly feature mining on the patch features of anomaly image to construct the anomaly set $\mathcal{M}'_{a}$. During the training stage, $\mathcal{M}'_{a}$ is augmented using a linear mixing strategy to obtain $\mathcal{M}_{a}^{*}$, and we train discriminators using $\mathcal{M}_{n}$ and $\mathcal{M}_{a}^{*}$ for anomaly classification. During the testing stage, we discard $\mathcal{M}_{n}$ and $\mathcal{M}_{a}^{*}$, and only utilize the discriminator for anomaly classification, generating an anomaly map and an anomaly score.}
   \label{fig:fig1}
\end{figure*}

\begin{itemize}
\item We introduce WSCXR, a weakly supervised anomaly detection framework for CXR, which effectively leverages anomaly features mining from few-shot anomaly images to enhance anomaly detection performance.
\item We propose a linear mixing strategy for augmenting anomaly features, addressing the significant quantity disparity between normal and anomaly features to enhance the generalization capability of the anomaly detectors.  
\item We conducted experiments on two CXR datasets, including ZhangLab \cite{kermany2018identifying} and CheXpert \cite{irvin2019chexpert}. The results demonstrate that WSCXR achieves state-of-the-art anomaly detection performance on both datasets.\footnote{The code is available at https://github.com/IamCuriosity/WSCXR.}
\end{itemize}

\section{Proposed Method}

In this section, we first present the problem definition of weakly supervised anomaly detection. Then, we introduce the workflow of WSCXR, which consists of three key steps: anomaly feature mining, linear mixing strategy, and discriminator training. The overall pipeline of WSCXR is illustrated in Fig. \ref{fig:fig1}.

\subsection{Problem Definition}

We denote the normal image set as $\mathcal{X}_{n}=\{(x_{i},y_{i}) \mid y_{i}=0, i=1,...,N\}$, and the anomaly image set as $\mathcal{X}_{a}=\{(x_{i},y_{i}) \mid y_{i}=1, i=1,...,K\}$, where $K<<N$ and $y_{i}$ represents the image-level labels. The training set $\mathcal{X}_{train}$ is defined as: $\mathcal{X}_{train} = \mathcal{X}_{n} \cup \mathcal{X}_{a}$. For the weakly supervised anomaly detection task, we train a detection model $\mathcal{F}(\cdot)$ using $\mathcal{X}_{train}$, so that for a given CXR image we have $\mathcal{A},\hat{y} = \mathcal{F}(x)$, where $\mathcal{A}$ represents the pixel-level anomaly map, and $\hat{y}$ the image-level anomaly score.

\begin{table*}[t]
  \renewcommand\arraystretch{1.4}
  \centering
  \large
  \caption{Comparison of WSCXR with alternative anomaly detection methods on the ZhangLab \cite{kermany2018identifying} and CheXpert \cite{irvin2019chexpert} datasets.}
   \resizebox{1.0\linewidth}{!}{
      \begin{tabular}{c|c|cccc|cc|cc|cc|cc}
    \bottomrule
    \multicolumn{2}{c|}{\multirow{3}[6]{*}{}} & \multicolumn{4}{c|}{\multirow{2}{*}{Unsupervised}} & \multicolumn{8}{c}{Weakly supervised} \\
\cline{7-14}    \multicolumn{2}{c|}{} & \multicolumn{4}{c|}{}         & \multicolumn{2}{c|}{K=2} & \multicolumn{2}{c|}{K=4} & \multicolumn{2}{c|}{K=8} & \multicolumn{2}{c}{K=16} \\
\cline{3-14}    \multicolumn{2}{c|}{} & CutPaste \cite{li2021cutpaste} & PatchCore \cite{roth2022towards} & SimpleNet \cite{liu2023simplenet} & SQUID \cite{xiang2023squid} & DDAD \cite{cai2022dual} & WSCXR & DDAD \cite{cai2022dual} & WSCXR & DDAD \cite{cai2022dual} & WSCXR & DDAD \cite{cai2022dual} & WSCXR \\
    \hline
    \multicolumn{1}{c|}{\multirow{3}[2]{*}{ZhangLab \cite{kermany2018identifying}}} & AUROC & 73.6$\pm$3.9 & 90.4$\pm$0.2 & 90.1$\pm$3.1 & 87.6$\pm$1.5 & 68.5$\pm$6.0 & 94.7$\pm$1.5 & 65.8$\pm$5.1 & 96.3$\pm$0.5 & 72.3$\pm$5.1 & 97.0$\pm$0.4 & 75.9$\pm$5.6 & \textbf{97.2$\pm$0.2} \\
          & ACC   & 64.0$\pm$6.5 & 81.8$\pm$1.6 & 84.5$\pm$2.4 & 80.3$\pm$1.3 & 68.0$\pm$0.8 & 90.0$\pm$2.2 & 66.9$\pm$3.5 & 91.4$\pm$0.3 & 67.1$\pm$3.0 & 92.3$\pm$0.7 & 70.0$\pm$4.6 & \textbf{92.7$\pm$0.3} \\
          & F1    & 72.3$\pm$8.9 & 84.1$\pm$1.9 & 86.9$\pm$2.4 & 84.7$\pm$0.8 & 74.6$\pm$3.2 & 91.8$\pm$1.8 & 73.7$\pm$3.0 & 93.1$\pm$0.2 & 71.4$\pm$3.4 & 93.8$\pm$0.7 & 74.4$\pm$4.8 & \textbf{94.0$\pm$0.4} \\
    \hline
    \multicolumn{1}{c|}{\multirow{3}[2]{*}{CheXpert \cite{irvin2019chexpert}}} & AUROC & 65.5$\pm$2.2 & 72.1$\pm$0.3 & 71.4$\pm$3.0 & 78.1$\pm$5.1 & 66.7$\pm$1.7 & 79.3$\pm$1.0 & 65.6$\pm$1.3 & 82.2$\pm$1.8 & 67.1$\pm$0.7 & 82.4$\pm$0.8 & 67.7$\pm$1.4 & \textbf{84.0$\pm$0.3} \\
          & ACC   & 62.7$\pm$2.0 & 68.5$\pm$0.6 & 66.7$\pm$1.8 & 71.9$\pm$3.8 & 64.2$\pm$2.1 & 73.3$\pm$0.6 & 62.8$\pm$0.7 & 76.3$\pm$2.0 & 64.0$\pm$0.8 & 76.7$\pm$0.8 & 64.5$\pm$1.4 & \textbf{78.7$\pm$0.5} \\
          & F1    & 60.3$\pm$4.6 & 69.2$\pm$1.1 & 66.6$\pm$2.8 & 75.9$\pm$5.7 & 62.0$\pm$4.1 & 74.1$\pm$1.0 & 58.7$\pm$6.6 & 77.0$\pm$1.4 & 58.0$\pm$4.9 & 78.1$\pm$0.9 & 63.0$\pm$2.1 & \textbf{78.9$\pm$0.6} \\
    \toprule
    \end{tabular}%
    }
  \label{tab:table1}
\end{table*}

\subsection{Anomaly Feature Mining}

For a normal image $x \in \mathcal{X}_{n}$, we first extract the features $\phi_j(x) \in R^{H_j \times W_j\times C_j}$ from a pretrained network $\phi$, where $\phi_j(x)$ represents the $j$th layer's feature, $H_j$, $W_j$, and $C_j$ respectively represent the height, width, and number of channels of the $j$th layer's features. For a given image $x$, we can obtain a set of pre-trained multi-scale features $\{\phi_1(x),...,\phi_j(x)\}$. In CXR images, lesion regions typically have a specific scale range; hence, we select a multi-scale pre-trained feature subset for anomaly detection. We upsample the features in the selected subset to the same spatial resolution and perform channel-wise alignment to obtain the aligned multi-scale feature $\phi(x) \in R^{H' \times W' \times C'}$. For a spatial position $(h,w)$, we denote $\phi_{(h,w)}(x) \in R^{C'}$ as the patch feature of $\phi(x)$ at position $(h,w)$. We follow \cite{roth2022towards} to aggregate features from neighboring regions, enhancing the spatial receptive field of patch feature, ensuring that the lesion regions can be detected comprehensively and continuously. For $\phi_{(h,w)}(x)$, the aggregated patch feature $\mathcal{Z}_{(h,w)}^{p}(x)$ is defined as follows:

\begin{equation}
\label{eq:equ1}
\mathcal{Z}_{(h,w)}^{p}(x) = f_{agg}(\{\phi_{(a,b)}(x) \mid (a,b) \in \mathcal{N}_{(h,w)}^{p}\})
\end{equation}
where $p$ represents the size of the aggregation neighborhood, $f_{agg}$ is an aggregation function, and $\mathcal{N}_{(h,w)}^{p}$ represents the set of aggregation positions:
\begin{gather}
\label{eq:equ2}
\begin{align}
\mathcal{N}_{(h,w)}^{p} = \{(a,b) \mid &a \in [h-\lfloor p/2 \rfloor,h+ \lfloor p/2 \rfloor], \notag \\
                                       &b \in [w-\lfloor p/2 \rfloor,w+\lfloor p/2 \rfloor]\}
\end{align}
\end{gather}
Next, we denote the feature set of image $x$ as $\mathcal{M}(x)$:
\begin{equation}
\label{eq:equ3}
\mathcal{M}(x) = \{ \mathcal{Z}_{(h,w)}^{p}(x) \mid h<H', w<W', h,w \in \mathbb{N} \}
\end{equation}
Given the normal image set $\mathcal{X}_{n}$, we can obtain the normal features set $\mathcal{M}_{n}= \bigcup_{x \in \mathcal{X}_{n}} \mathcal{M}(x)$. Likewise, the set of anomaly image features is $\mathcal{M}_{a}= \bigcup_{x \in \mathcal{X}_{a}} \mathcal{M}(x)$. Due to the presence of a significant number of normal regions in the anomaly images, the anomaly feature set $\mathcal{M}_{a}$ contains a considerable amount of normal patch features. We employ a anomaly feature mining strategy to eliminate normal patch features from $\mathcal{M}_{a}$, resulting in the set $\mathcal{M}'_{a}$, which exclusively comprises anomaly patch features, with $\mathcal{M}'_{a} \subseteq \mathcal{M}_{a}$.

For $\forall m \in \mathcal{M}_{a}$, we define the function $\mathcal{S}(m)$ to assess the degree of anomaly in patch feature $m$:
\begin{equation}
\label{eq:equ4}
\mathcal{S}(m) = \min_{m_n \in \mathcal{M}_{n}} \| m_n-m \|_{2}
\end{equation}
$\mathcal{S}(m)$ retrieves the feature in $\mathcal{M}_{n}$ that is closest to $m$, using the \texttt{L2} distance between them as the anomaly score for $m$. Then, we select the Top$K$ features with the largest $\mathcal{S}(m)$ in $\mathcal{M}_{a}$ as the set $\mathcal{M}'_{a}$:
\begin{equation}
\label{eq:equ5}
\mathcal{M}'_{a} = \mathop{\arg\max}\limits_{\mathcal{M} \subseteq \mathcal{M}_{a}} \sum_{m \in \mathcal{M}}\mathcal{S}(m)
\end{equation}
The size of $\mathcal{M}'_{a}$ is determined by the retention rate $r$ ($0 < r \leq 1$), and $\lvert \mathcal{M}'_{a} \rvert = \lfloor r\times \lvert \mathcal{M}_{a} \rvert \rfloor$. The features in $\mathcal{M}'_{a}$ exhibit the largest \texttt{L2} distance from $\mathcal{M}_{n}$. We consider this to indicate a significant dissimilarity between the regions corresponding to $\mathcal{M}'_{a}$ and the normal regions of the image. This demonstrates that $\mathcal{M}'_{a}$ is a mined set of lesion regions features.

\subsection{Linear Mixing Strategy}

Due to the significant disparity in the quantities between normal and anomaly images, resulting in $\lvert \mathcal{M}_{n} \rvert >> \lvert \mathcal{M}'_{a} \rvert$, using $\mathcal{M}_{n}$ and $\mathcal{M}'_{a}$ for training of anomaly detector would lead to severe overfitting, thereby degrading the performance of anomaly detection. To address this issue, we propose a linear mixing strategy for augmenting anomaly features. Specifically, for $\forall m_n \in \mathcal{M}_{n}$ and $\forall m_a \in \mathcal{M}'_{a}$, we perform linear mixing to obtain the new anomaly feature $m'_a$:
\begin{equation}
\label{eq:equ6}
m'_a =\alpha m_a + (1-\alpha) m_n 
\end{equation}
where $\alpha$ ($0 < \alpha \leq 1$) is the mixing factor that determines the degree of anomaly in $m'_a$. We use the $m'_a$ to augment the anomaly set to obtain the augmented anomaly feature set $\mathcal{M}_{a}^{*}$, resulting in $\lvert \mathcal{M}_{n} \rvert \approx \lvert \mathcal{M}_{a}^{*} \rvert$.

\subsection{Discriminator Training}

We use $\mathcal{M}_{n}$ and $\mathcal{M}_{a}^{*}$ to train the discriminator $\mathcal{D}$, with the following loss function $\mathcal{L}$:

\begin{gather}
\label{eq:equ7}
\begin{align}
\mathcal{L} = \frac{1}{\lvert \mathcal{M}_{n} \rvert  \lvert \mathcal{M}_{a}^{*} \rvert}\sum_{m_n \in \mathcal{M}_{n}}\sum_{m_a \in \mathcal{M}_{a}^{*}} & max[0,\mathcal{D}(m_n)]  \notag \\
+ &max[0,1-\mathcal{D}(m_a)]          
\end{align}
\end{gather}
During the test phase, we discard $\mathcal{M}_{n}$ and $\mathcal{M}_{a}^{*}$ using only the discriminator to generate anomaly scores. For a test image $x_t $, we can obtain the anomaly map $\mathcal{A}(x_t)$ of $x_t$ as:
\begin{equation}
\label{eq:equ8}
\mathcal{A}(x_t) = \{ \mathcal{D}(m_t) \mid m_t \in \mathcal{M}(x_t)\}
\end{equation}
We use the maximum value in $\mathcal{A}(x_t)$ as the image-level anomaly score $\hat{y}$ of $x_t$.

\section{Experiments}

\subsection{Experimental Setup}

\textbf{Datasets.} We conducted comprehensive experiments on two CXR datasets, including the ZhangLab Chest X-Ray dataset \cite{kermany2018identifying} and the Stanford CheXpert dataset \cite{irvin2019chexpert}. The ZhangLab dataset \cite{kermany2018identifying} comprises healthy and pneumonia CXR images from pediatric patients aged from one to five years, with pneumonia images serving as anomaly images. The CXR images in the CheXpert dataset \cite{irvin2019chexpert} are sourced from clinical patients at Stanford Hospital and encompass 12 different diseases (such as Cardiomegaly, Edema, Pleural Effusion, Pneumonia, etc.). We followed the settings in \cite{xiang2023squid}. For the ZhangLab dataset \cite{kermany2018identifying}, we used 1,349 normal images during the training phase, and the test set consisted of 234 normal images and 390 abnormal images. For the CheXpert dataset \cite{irvin2019chexpert}, during the training phase, we utilized 4,500 normal images, and the test set included 250 normal images and 250 anomaly images. Additionally, for each anomaly type, we provided an additional $K \in \{2, 4, 8, 16\}$ anomaly images for WSCXR during the model training.

\textbf{Implementation Details.} We used a DenseNet121 \cite{huang2017densely} pre-trained on ImageNet \cite{deng2009imagenet} for feature extraction, retaining the features from the fourth layer for anomaly detection. For the aggregation neighborhood size $p$ in Eq. \eqref{eq:equ1}, we set it to 5. And we used adaptive average pooling as the aggregation function $f_{agg}$ in Eq. \eqref{eq:equ1}. In the anomaly feature mining process, we set the retention rate $r$ to 0.2. The mixing factor in Eq. \eqref{eq:equ6} was uniformly sampled in [0.1, 1.0]. The discriminator was implemented with a basic MLP, mapping patch features to anomaly scores. All experiments were independently evaluated three times.

\textbf{Baselines and Metrics.} To comprehensively evaluate the anomaly detection performance of WSCXR, we established the following baselines for comparison: For unsupervised anomaly detection methods (using only normal images during the training phase), we compared CutPaste (CVPR'21) \cite{li2021cutpaste}, PatchCore (CVPR'22) \cite{roth2022towards}, SimpleNet (CVPR'23) \cite{liu2023simplenet}, and SQUID (CVPR'23) \cite{xiang2023squid}. For weakly supervised anomaly detection methods, we compared DDAD (MICCAI'22) \cite{cai2022dual}, where we combined normal image set and anomaly image set as the unlabeled image set in \cite{cai2022dual}. For the evaluation metrics, we followed the settings in \cite{xiang2023squid}, using image-level AUROC, Accuracy, and F1 scores.

\begin{table}[t]
  \renewcommand\arraystretch{1.3}
  \centering
  \caption{Ablation studies of WSCXR on the ZhangLab dataset \cite{bergmann2019mvtec} with $K=16$.}
   \resizebox{1.0\linewidth}{!}{
    \begin{tabular}{c|cc|ccc}
    \bottomrule
          & \makecell[c]{Anomaly \\ Feature Mining} & Linear Mixing & AUROC & ACC & F1 \\
    \hline
    1     & -     & -     & 89.2$\pm$1.5 & 83.1$\pm$2.3 & 84.8$\pm$0.4 \\
    \hline
    2     & \checkmark     & -     & 94.9$\pm$0.9 & 89.2$\pm$0.5 & 91.0$\pm$0.4 \\
    \hline
    3     & -     & \checkmark    & 92.2$\pm$1.5 & 85.2$\pm$2.0 & 86.9$\pm$1.8 \\
    \hline
    4     & \checkmark     & \checkmark     & \textbf{97.2$\pm$0.2} & \textbf{92.7$\pm$0.3} & \textbf{94.0$\pm$0.4} \\
    \toprule
    \end{tabular}%

    }
  \label{tab:table2}
\end{table}

\subsection{Experimental Results}

We compared WSCXR with alternative anomaly detection methods on the ZhangLab \cite{kermany2018identifying} and CheXpert \cite{irvin2019chexpert} datasets, and the experimental results are presented in Table \ref{tab:table1}. Compared to other unsupervised anomaly detection methods, WSCXR demonstrates a significant performance improvement using just two anomaly images. As the number of provided anomaly images increases, WSCXR's performance continues to improve, indicating its effective utilization of crucial visual cues within anomaly images. In comparison to DDAD \cite{cai2022dual}, WSCXR achieves a notable performance gain, achieving a 20\% and 10\% performance improvement on the ZhangLab \cite{kermany2018identifying} and CheXpert \cite{irvin2019chexpert} datasets respectively.

\subsection{Ablation Studies}

We validated the impact of anomaly feature mining and linear mixing strategy on the anomaly detection performance of WSCXR using the ZhangLab \cite{kermany2018identifying} dataset, with the number of anomaly images set to 16, as shown in Table \ref{tab:table2}. In experiments 1 and 3, we excluded anomaly feature mining and used the anomaly feature set $\mathcal{M}_{a}$ for subsequent discriminator training or linear mixing. In experiments 1 and 2, we excluded linear mixing strategy and used $\mathcal{M}_{a}$ or $\mathcal{M}'_{a}$ for discriminator training. The experimental results indicate that anomaly feature mining contributes significantly to performance improvement. The original anomaly feature set $\mathcal{M}_{a}$ contains a substantial number of normal region features. Training with $\mathcal{M}_{a}$ would confuse the discriminator, leading to a significant decrease in anomaly detection performance. Furthermore, the use of the linear mixing strategy to augment the anomaly feature set effectively eliminates the difference in the quantity between the normal feature set and anomaly feature set. This enables the discriminator to learn a more compact decision boundary, thereby effectively improving the anomaly detection performance of WSCXR.

\begin{figure}[t]
  \centering
   \includegraphics[width=\linewidth]{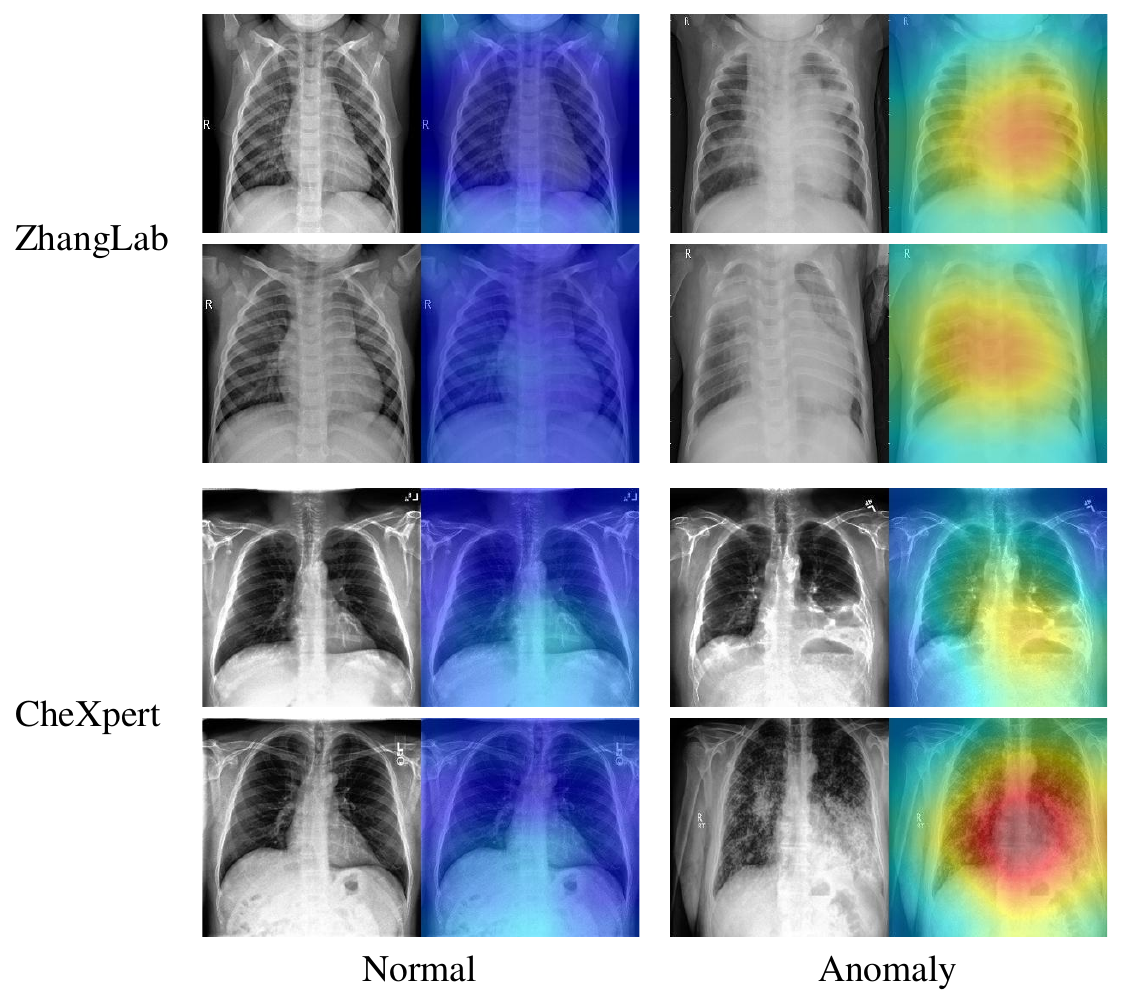}
   \caption{Visualization results of WSCXR on the ZhangLab \cite{kermany2018identifying} and CheXpert \cite{irvin2019chexpert} datasets. Within each group, from left to right, are the CXR image and predicted anomaly map.}
   \label{fig:fig2}
\end{figure}

\subsection{Visualization}

To validate the lesion localization performance of WSCXR, we provide the visual analysis on the ZhangLab \cite{kermany2018identifying} and CheXpert \cite{irvin2019chexpert} datasets, and the experimental results are shown in Fig. \ref{fig:fig2}. For each group, we provide the original CXR images and the predicted anomaly map. The qualitative results demonstrate that WSCXR can accurately distinguish between normal and anomaly images and precisely localize the lesion regions in anomaly CXR images. This provides valuable references for medical decision-making in real life.

\section{Conclusion}

This letter investigates the problem of medical image anomaly detection under weakly supervised settings, an important but underexplored field. In this respect, we propose WSCXR, a weakly supervised Chest X-Ray anomaly detection framework. WSCXR effectively leverages crucial visual cues in few-shot anomaly images for anomaly detection. Compared to traditional unsupervised anomaly detection methods, WSCXR jointly models normal and anomaly images, enabling the model to learn more compact decision boundaries. We conducted experiments on two Chest X-Ray datasets, demonstrating that WSCXR surpasses existing state-of-the-art methods with just two anomaly images. Furthermore, we validated the effectiveness of each module of WSCXR through ablation studies and provided the visual analysis. Our experiments suggest that WSCXR has the potential to assist medical professionals in making clinical decisions.

\bibliographystyle{IEEEtran}
\bibliography{bare_jrnl-docx}

\end{document}